\documentclass[sigconf,screen]{acmart}
\usepackage{xcolor}
\usepackage{wrapfig}

\usepackage{amsmath}
\usepackage{algorithm}
\usepackage{algpseudocode}

\usepackage{multirow}
\usepackage[normalem]{ulem}
\useunder{\uline}{\ul}{}
\usepackage{diagbox}

\usepackage{pifont}

\usepackage{bbding}

\usepackage{hhline}
\usepackage{colortbl}
\usepackage{framed}
\usepackage{mdframed}
\usepackage{amsmath}
\usepackage{array}
\usepackage[caption=false]{subfig}
\usepackage{listings}
\definecolor{darkred}{rgb}{0.6,0.0,0.0}

\definecolor{mycolor1}{RGB}{0, 0, 180}
\definecolor{mycolor2}{RGB}{31, 132, 31}
\usepackage[true]{anonymous-acm}

\usepackage[compact]{titlesec}

\usepackage{microtype}

\algnewcommand\algorithmicoutput{\textbf{Output:}}
\algnewcommand\OUTPUT{\item[\algorithmicoutput]}

\algnewcommand\algorithmicdefine{\textbf{Define:}}
\algnewcommand\DEFINE{\item[\algorithmicdefine]}

\vspace{-10pt}

\def\BibTeX{{\rm B\kern-.05em{\sc i\kern-.025em b}\kern-.08em
    T\kern-.1667em\lower.7ex\hbox{E}\kern-.125emX}}
    
\setcopyright{rightsretained}
\acmDOI{10.1145/3663529.3663779}
\acmYear{2024}
\copyrightyear{2024}
\acmSubmissionID{fsecomp24ivr-p36-p}
\acmISBN{979-8-4007-0658-5/24/07}
\acmConference[FSE Companion '24]{Companion Proceedings of the 32nd ACM International Conference on the Foundations of Software Engineering}{July 15--19, 2024}{Porto de Galinhas, Brazil}
\acmBooktitle{Companion Proceedings of the 32nd ACM International Conference on the Foundations of Software Engineering (FSE Companion '24), July 15--19, 2024, Porto de Galinhas, Brazil}
\received{18-JAN-2024}
\received[accepted]{2024-04-09}

\begin{document}

\title{Testing Learning-Enabled Cyber-Physical Systems with Large-Language Models: A Formal Approach}


\author{Xi Zheng}
\affiliation{
  \institution{Macquarie University}
  \country{Australia}
}
\email{james.zheng@mq.edu.au}

\author{Aloysius K. Mok}
\affiliation{
    \institution{University of Texas at Austin}
    \country{USA}
}
\email{mok@cs.utexas.edu}

\author{Ruzica Piskac}
\affiliation{
    \institution{Yale University}
   \country{USA}
}
\email{ruzica.piskac@yale.edu}

\author{Yong Jae Lee}
\affiliation{
    \institution{University of Wisconsin Madison}
    \country{USA}
}
\email{yongjaelee@cs.wisc.edu}


\author{Bhaskar Krishnamachari}
\affiliation{
    \institution{University of Southern California}
   \country{USA}
}
\email{bkrishna@usc.edu}

\author{Dakai Zhu}
\affiliation{
    \institution{University of Texas at San Antonio}
   \country{USA}
}
\email{Dakai.Zhu@utsa.edu}

\author{Oleg Sokolsky}
\affiliation{
    \institution{University of Pennsylvania}
    \country{USA}
}
\email{sokolsky@cis.upenn.edu}

\author{Insup Lee}
\affiliation{
    \institution{University of Pennsylvania}
    \country{USA}
}
\email{lee@seas.upenn.edu}



\renewcommand{\shortauthors}{Zheng et al.}

\begin{abstract}
The integration of machine learning into cyber-physical systems (CPS) promises enhanced efficiency and autonomous capabilities, revolutionizing fields like autonomous vehicles and telemedicine. This evolution necessitates a shift in the software development life cycle, where data and learning are pivotal. Traditional verification and validation methods are inadequate for these AI-driven systems. This study focuses on the challenges in ensuring safety in learning-enabled CPS. It emphasizes the role of testing as a primary method for verification and validation, critiques current methodologies, and advocates for a more rigorous approach to assure formal safety.
\end{abstract}




\maketitle
\pagestyle{plain}

\section{CONTEXT, MOTIVATIONS AND AIMS}
The integration of machine learning (ML) with cyber-physical systems (CPS) has revolutionized various sectors, including transportation, logistics, service industries, and healthcare, with innovations like autonomous vehicles (Waymo~\cite{Waymo}, Tesla Autopilot~\cite{Tesla}, Uber ATG~\cite{Uber}), delivery drones (Amazon Prime Air~\cite{AmazonPrimeAir}, Google Wing~\cite{GoogleWing}, Zipline~\cite{Zipline}), and robotic surgeries (Da Vinci~\cite{DaVinci}, Mazor~\cite{Mazor}, Mako~\cite{Mako}). However, these advancements have raised significant safety concerns, evidenced by reported incidents causing fatalities and economic loss~\cite{CarCrashes, WingFailure,RoboticsArmFailure,singh2016instrument}. 

Addressing these issues requires rigorous verification and validation, presenting unique challenges due to the complexities of learning-enabled CPS. These systems incorporate critical machine learning components like perception and planning, as seen in autonomous driving, making them significantly different from traditional software systems.
This complexity, involving a paradigm shift in the software development life cycle to incorporate data and learning, demands novel approaches in verification and validation techniques~\cite{sommerville2011software, amershi2019software}.

We present our initial efforts to explore practical testing strategies for the verification and validation of learning-enabled CPS. This focus is particularly relevant given the extensive use of testing in the CPS industry and the considerable amount of recent literature on this topic. From a summary of the current state-of-the-art testing methodologies for learning-enabled CPS, we propose a roadmap to formalize the testing effort.

More specifically, we use large-language models (LLMs) to extract human knowledge from existing rules and regulations, and analyze vast amounts of data generated or captured by learning-enabled CPS, including sensor data and logs. By extracting human knowledge and analyzing data, LLMs can offer insights into the system's behavior and generate a wealth of realistic and high-quality test data. With this improved data quality, it becomes feasible to employ data-driven learning to extract underlying formal specifications. The large language models can also be tasked with identifying corresponding formal specifications. Based on these derived formal specifications, we can engage in model-based testing—a considerably more formal approach than existing methodologies, such as search-based testing.

We present a case study utilizing a vision-based LLM to analyze traffic accident from photos, showing promising results. This lays a foundation for employing multi-modal LLMs to distill meaningful latent representations from real-world sensor data in learning-enabled CPS. For instance, leveraging front-facing cameras in autonomous vehicles to capture diverse traffic incidents. Another study demonstrates using GPT-4 and a customized domain-specific language to extract knowledge and generate test scenarios from a real-world traffic handbook. Early results indicate potential in uncovering diverse bugs in autonomous driving systems, aligning well with the roadmap outlined for formal testing and verification.

\section{RELATED WORK}
\label{sec:relatedwork}
Recent research in formal verification of machine learning models and testing of learning-enabled CPS has used various techniques like NNV star sets~\cite{tran2020verification}, Sherlock~\cite{dutta2019sherlock}, Reluplex~\cite{katz2017reluplex}, and Branch and Bound~\cite{bunel2018unified}. However, these approaches often lack in providing comprehensive safety guarantees and are limited in handling the complexity of industry-scale, multi-modal CPS~\cite{lou2022testing, prakash2021multi}. Similarly, while testing in system properties ~\cite{deng2022declarative, tian2022mosat, li2020av} and system robustness~\cite{tian2018deeptest, deng2020analysis, cai2020real} are progressing, they fail to offer concrete formal assurances. This highlights the gap in current methodologies and underscores the need for more robust testing approaches that can ensure safety and reliability in real-world applications of CPS~\cite{shalev2017formal, seshia2017compositional, zapridou2020runtime}.

\section{A ROADMAP TOWARDS FORMAL TESTING}
\label{sec:roadmap}
\begin{figure*}[tb!]
\centering
\includegraphics[width = 1\textwidth]
{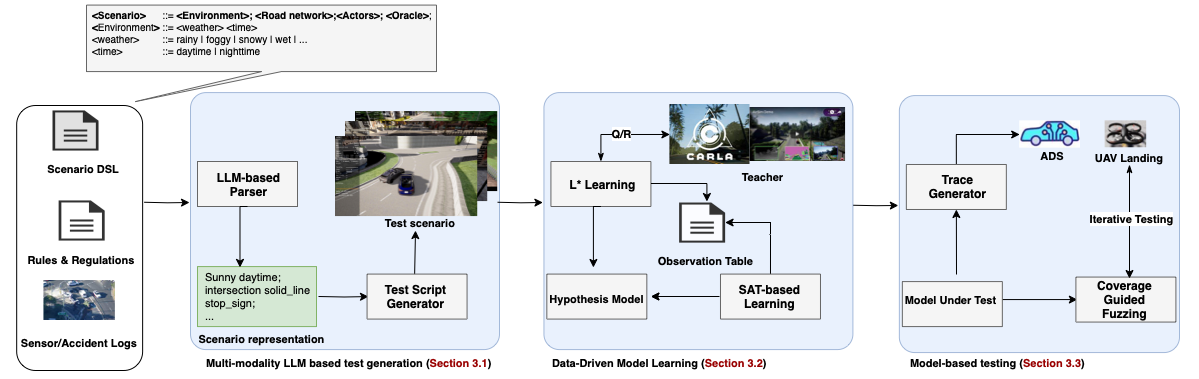}
\caption{Proposed Roadmap}
\label{fig:roadmap}
\end{figure*}

Our roadmap, depicted in Fig.\ref{fig:roadmap}, targets Multi-Modal LLM-based Test Generation, Data-Driven Model Learning, and Model-based Testing. We begin by capturing test scenarios from varied sources such as rules, sensor data, and accident logs, processing them with domain-specific languages like OpenScenario~\cite{openScenario} and Scenic~\cite{fremont2022scenic} via an LLM-based parser (ChatGPT or LLaVA~\cite{liu2023visual}). This initial phase is vital for generating diverse test cases, particularly for safety and liveness properties. Subsequent phases involve Data-Driven Model Learning using the $L^*$ algorithm for model construction and validation through high-fidelity simulations in collaboration with industry leaders~\cite{deng2020rmt,deng2022scenario,deng2023target}. The final phase, Model-based Testing, utilises these models to generate test cases that formally assure compliance with specified requirements. This approach is designed to effectively bridge the gap between model checking and runtime verification in learning-enabled CPS.

\subsection{Multi-modal LLM based Test Generation}
\label{sec:MMLLMTestGen}
Generating diverse test cases that violate the formal properties of learning-enabled CPS, setting the conditions for our second stage, is a significant hurdle~\cite{lou2022testing}. This complexity arises from a vast search space~\cite{tian2022mosat, li2020av, deng2020analysis} and limited domain knowledge~\cite{tian2018deeptest, deng2022declarative}. 
On the other hand, LLMs are being intensively studied for their applications in personalized learning~\cite{kasneci2023chatgpt}, software testing~\cite{wang2023software}, and multi-modal scenarios~\cite{huang2023chatgpt}.
In learning-enabled CPS, abundant data including system logs and sensor outputs (such as camera and lidar feeds) offer opportunities for fine-tuning multi-modal LLMs. These refined models can interpret the semantics of various physical environments and their interaction with the system. For example, in some countries, front-facing cameras are widely installed in vehicles to deter ``pedestrian scams'', where individuals deliberately throw themselves in front of cars to extort drivers or launch false insurance claims~\cite{button2016shallow}. An ancillary benefit of this is that multi-angle video data can be collected for individual traffic incidents. Such data, coupled with accident descriptions, can be used to train video-based LLMs to understand the nuanced factors leading to accidents.

These fine-tuned LLMs can then facilitate test generation in various ways. For example, they can be employed for seed generation, mutation, and selection in fuzzing techniques, a concept already explored in traditional software domains~\cite{deng2023large,zhang2023understanding}. Similarly, in the case of autonomous drone systems, drones capture a range of images of surveying or landing sites~\cite{alam2021survey,dilshad2020applications}. These images, along with local government guidelines on surveying and landing, can be utilised to generate relevant test cases. 
To mitigate hallucination, we implement multi-stage validation to ensure the correctness of LLM outputs, as demonstrated in our pilot study briefly discussed in Section~\ref{sec:Target}.


\subsection{Data-driven Model Learning}
\label{sec:DDML}
Model learning techniques, from Biermann's offline approach~\cite{biermann1972synthesis} to Angluin's online $L^*$ algorithm~\cite{angluin1987learning}, have been thoroughly explored. Online methods generally outperform offline methods, being polynomial rather than NP-complete~\cite{leucker2006learning}. Implementing membership and equivalence queries, however, introduces practical challenges due to the complexities of real system tests~\cite{steffen2011introduction}. Recent studies suggest alternatives like falsification for equivalence queries~\cite{annpureddy2011s} and using positive examples for learning signal temporal logic (STL) properties~\cite{jha2017telex}, though these require extensive setup or strong assumptions~\cite{jones2014anomaly, bartocci2022survey}. Our goal is to refine system behavior modelling using structured learning to bypass these traditional automata-learning limitations, utilising diverse corner cases generated in the prior step and leveraging our experience in building high-fidelity simulation/co-simulation environments as oracles, we can effectively address the queries needed for the adapted $L^*$ algorithm.

The process begins with the $L^*$ learning algorithm, which issues membership queries to a simulator acting as an oracle for the underlying CPS, such as Carla~\cite{dosovitskiy2017carla} for autonomous driving and AirSim for unmanned aerial vehicles~\cite{airsim2017fsr}. These queries are instrumental in exploring the system's behavior across a myriad of scenarios generated by the LLM in Section~\ref{sec:MMLLMTestGen}, with the simulator's feedback aiding the construction of a preliminary hypothesis model.


The membership queries, categorized into positive and negative test cases executed and evaluated by the simulator, are pivotal in shaping the hypothesis model. Positive test cases provide insights into valid system behaviors, while negative test cases highlight invalid or erroneous behaviors. This dichotomy aids in the accurate refinement of the hypothesis model, steering it towards a more accurate representation of the system's dynamics. As the hypothesis model evolves, the role of equivalence queries becomes important. These queries are directed at the simulator to ascertain the consistency between the hypothesis model and the simulated system behavior. Discrepancies uncovered through equivalence queries point out areas where the hypothesis model needs further refinement.

Parallelly, the SAT-based learning algorithm similar to~\cite{neider2018learning} enhances this iterative learning process by managing the logical structuring of the constraints derived from both membership and equivalence queries. It reviews the consistency of these constraints against the hypothesis model, identifying and rectifying inconsistencies. This complementary role of SAT-based learning augments the $L^*$  learning, ensuring a more precise and robust hypothesis model. The synergy between $L^*$  learning, SAT-based learning, and the strategic deployment of equivalence queries, all interfaced with the simulator, crafts a robust framework for model extraction. This iterative, multi-faceted approach progressively refines the hypothesis model, aligning it closely with the simulated system behavior. The simulator's role as an oracle is crucial, providing a practical and reliable benchmark for validating and refining the hypothesis model.

In summary, this blended learning approach, rooted in iterative interaction with a simulator oracle, presents a viable pathway for extracting accurate models from complex systems. Through an iterative engagement of $L^*$ learning, SAT-based learning, and strategic querying, this proposed solution promises a substantial advancement in the field of model extraction, especially in scenarios where a concrete system specification is absent.

\subsection{Model-based Testing}
\label{sec:MBT}

In this paper, we aim to rejuvenate the concept of ``model-based testing'' within the realm of learning-enabled CPS. One of the earliest proponents of this concept hails from Bellcore, where a test data model was articulated using a straightforward specification named AETGSpec. This specification supports hierarchy in both fields and relations~\cite{dalal1999model}. Building on this foundation, Schiefer et al.~\cite{schieferdecker2012model} highlighted that test case generation for model-based testing can adopt various methods. One approach is deductive theorem proving, wherein the model is segmented into equivalence classes based on a set of logical expressions. In its most basic form, each class can function as a test case. In the context of model checking, test case generation revolves around identifying counterexamples where the specification is breached. Symbolic execution can be employed to navigate every potential program execution path. Tools like Modbat~\cite{artho2013modbat}, tailored for event-driven systems, and MoMuT~\cite{krenn2015momut}, designed for UML and timed automata, facilitate test case generation from state machine models.
However, such black-box type of approach relies on random and mutation tests are not applicable to learning-based CPS as key learning models are not applicable to random seed generation and mutation operators.

In our research, we gravitate towards generating test cases steered by STL fuzzing due to STL's ability to express complex temporal and spatial relationships within CPS. In a contemporary study by Meng et al.~\cite{meng2022linear}, given a linear-time temporal logic (LTL) property $\phi$, a B\"uchi automaton $\mathrm{A}\neg\phi$ can be crafted that recognizes the contravention of the property $\phi$. 
 This approach requires manually extracting LTL properties, dealing with LTL's limited expressiveness compared to STL, and identifying program parts impacted by LTL's atomic propositions. The goal is to generate logs to identify potential proposition violations, guiding future test generation. This ensures focus on areas likely to breach LTL properties, enhancing test effectiveness. However, in learning-based CPS, like in autonomous vehicles where speed control is distributed across multiple neural networks, pinpointing specific locations for such violations is unfeasible, complicating the application of this strategy.

We propose using data-driven specifications from Section~\ref{sec:MMLLMTestGen} to identify STL properties and construct corresponding automata with suitable acceptance conditions for the negation of these properties. As learning-based CPS often use robotic operating systems with message queues, we will create a trace generator to monitor and record events for each STL predicate, generating relevant traces. Coverage-guided fuzzing will then be employed to produce test cases that activate trace data across sequential states up to the accepting state, creating counterexamples. Considering the complexity of the resulting timed automata, traversal methods such as depth-first, breadth-first, or random walk will be considered to ensure measurable coverage and formal assurance.


In conclusion, our model-based testing paradigm presents a significant improvement from existing fuzzing techniques. Where conventional coverage-guided fuzzers like AFL~\cite{wong2022american} primarily detect crashes and memory overflows, and fitness function-guided fuzzers~\cite{
haq2022efficient,haq2023many,zhong2022neural,li2020av} are designed for specific scenarios and oracles with no formal coverage guarantees, our approach elevates the discourse. Recent efforts parallel to ours, albeit requiring manual, error-prone specification of detailed scenarios and properties using new DSLs, underscore the laborious nature of these tasks~\cite{zhou2023specification}. Our model-based testing, on the other hand, automates scenario generation and property extraction, rendering it more accessible for industry practitioners dealing with black-boxed critical components in learning-enabled systems. The trace generator plugin in our framework is designed to cater to the extracted specifications, and our coverage-guided fuzzing not only maximizes failure coverage but also explores diverse counterexamples violating the tested property. This  approach, therefore, not only aligns with real-world industrial contexts but also opens new opportunities  in ensuring more robust, formally verified learning-enabled CPS.

\section{EARLY RESULTS AND EFFORTS}
\label{sec:earlyresults}
We conducted two case studies to investigate Research Question 1 (RQ1): assessing the ability of LLM to generate diverse yet real test cases through in-context learning. Similarly, initial evaluations were made for Research Question 2 (RQ2): examining a multi-modal LLM's understanding of traffic accident's root cause. We will share some promising initial results, aligning well with our roadmap.

\subsection{RQ1: Test Case Generation Capability using LLM}
\label{sec:Target}
Using ChatGpt4.0, we interpreted the Texas traffic rule handbook to create a DSL (Fig.~\ref{fig:schema}) that transforms these rules into traffic scenarios. This DSL, focusing on semantic descriptions rather than precise coordinates, differentiates from others such as OpenScenario and Scenic~\cite{openScenario, geoscenario, Scenic}. 
We implemented multi-level validation to ensure the correctness of the DSL specifications, mitigating hallucination issues.
 We then translated these DSL specifications into test scripts for the CARLA simulation platform~\cite{dosovitskiy2017carla}, uncovering significant bugs in autonomous driving systems. The DSL comprises elements like \textit{Environment}, \textit{Road network}, \textit{Actor}, and \textit{Oracle}, each capturing intricate scenario semantics. This approach enabled us to generate diverse, semantically rich test scenarios that revealed rule violations in real-world autonomous systems, aiding developers in pinpointing specific issues. In our experiments, we observed that the MMFN model~\cite{zhang2022mmfn} failed to stop at the stop sign, Autoware~\cite{kato2018autoware} collided with a front vehicle, and LAV~\cite{chen2022lav} did not respond to a pedestrian crossing the road. We reported these issues, along with detailed log data, to the developers of these widely-used autonomous driving systems. They utilised our logs to pinpoint the root causes of these rule violations. 
Further details about the DSL, including insights, examples, and the methodology for translating DSLs into test scripts, along with replicable artifacts, are provided in~\cite{deng2023target}.

\begin{figure}[tb!]
\centering
\scalebox{0.95}{
{
$\def\arraystretch{1.1}
\setlength{\arraycolsep}{1pt}
\begin{array}{lll}
\textbf{\em <Scenario>} &::= \text{\em <Environment>}; \ \text{\em <Road network>}; \\ \ & \ \ \ \ \ \ \text{\em <Actors>}; \ \text{\em <Oracle>}; \\
\textbf{\em <Environment>} &::= \text{\em <weather>} \ \text{\em <time>} \\
\textbf{\em <Road network>} &::= \text{\em <road type>} \ \text{\em <road marker>} \ \text{\em <traffic signs>} \  \\
\textbf{\em <Actors>} &::= \text{\em <ego vehicle>}, \  \text{\em <npc actors>} \\
\textbf{\em <Oracle>} &::= \text{\em <longitudinal oracles>} \ \text{\em <lateral oracles>} \\
\text{...}
\end{array}$
}}
\caption{High-level structure of the Scenario DSL} 
\label{fig:schema}
\end{figure}

\subsection{RQ2: Accident Root Cause Analysis using Multi-Modal LLM}
\begin{wrapfigure}{R}{0.25\textwidth}
\begin{center}
\includegraphics[width=0.25\textwidth]{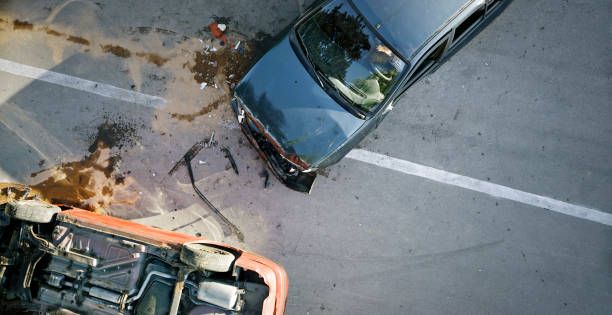}
\end{center}
\vspace{-0.5cm}
\caption{A single frame taken from a traffic collision video}
\label{fig:accident} 
\end{wrapfigure} 

In this study, we utilized a vision encoder-based multi-modal Large Language Model (LLaVA~\cite{liu2023visual}) to closely examine specific traffic accidents. Our objective is to investigate the boundaries of what such a multi-modal LLM can discern about the causes of these incidents. We hypothesize that if the LLM can accurately comprehend these causes, then its latent vector representation should effectively capture essential features. Consequently, this would enable the LLM to generate a variety of ``corner cases'' that closely resemble real-world traffic incidents. These test cases serve dual functions: they can evaluate the robustness of autonomous driving systems (Section~\ref{sec:MMLLMTestGen}) and also pave the way for our future work in data-driven learning (Section~\ref{sec:DDML}) and model-based testing (Section~\ref{sec:MBT}).

As demonstrated in Figure~\ref{fig:accident}, without fine-tuning or in-context learning, when we posed the question ``Can you describe the scene?'', LLaVA responded, ``The scene in the image shows a chaotic and damaged street, with two cars involved in a collision. One of the cars has been flipped over, and debris is scattered around the area. The accident has caused significant damage to both vehicles...''.

When we modified the query to ``Can you imagine what leads to such a collision?'', the LLM responded, listing several common causes including speeding, distracted driving, and poor visibility. The first response accurately depicts the accident, while the rest elaborate on potential contributing factors.

We are transitioning from an image-based to a video-based vision encoder in our LLM, expecting improved confidence and temporal information capture. This enhancement should better detect key features of traffic incidents and facilitate the creation of realistic and complex corner cases, as outlined in stage 1 of our plan.



\section{Conclusion}
\label{sec:conclusion}
This paper critically evaluates the existing formal verification processes for learning-enabled CPS and highlights the limitations of traditional software testing methods due to their lack of robust guarantees. Our forward-looking three-stage roadmap addresses key challenges, such as generating diverse corner cases, accessing sensor data ethically, and improving methods for timed automata extraction and state coverage. Our case studies demonstrate the roadmap's potential to fulfill the rigorous requirements of stakeholders like manufacturers, lawmakers, and customers. We anticipate this roadmap will catalyze collaborative efforts across communities to enhance formal guarantees in learning-enabled CPS.

\clearpage

\bibliographystyle{ACM-Reference-Format}
\bibliography{main}

\end{document}